# Non-local coherent screening in the bandwidth controlled $Ca_{1-x}Sr_xVO_3$ series

R.J.O. Mossanek[a], M. Abbate[a]*, A. Fujimori[b]

[a]*Departamento de Física, Universidade Federal do Paraná,*
*Caixa Postal 19091, 81531-990 Curitiba PR, Brazil*
[b]*Department of Physics, University of Tokyo, Bunkyo-ku, Tokyo 113, Japan*

We studied the electronic structure of the $Ca_{1-x}Sr_xVO_3$ series using a cluster model. Both the *coherent* and *incoherent* structures in photoemission correspond to *well screened* states. The *coherent* structure ($3d^1\underline{C}$) is mostly due to a non-local *coherent screening*, whereas the *incoherent* feature ($3d^1\underline{L}$) is mainly related to the *ligand screening*. The *poorly screened* state ($3d^0$) forms the lower Hubbard band and appears deeper in energy. The increased distortion in the series decrease the V-O-V angle and thus the non-local transfer integral T*. This produces the observed transfer of spectral weight from the *coherent* to the *incoherent* structure. A further decrease of T* would eventually induce a bandwidth controlled metal-insulator transition.

Highly correlated systems present strong Coulomb correlations U and relatively small bandwidth W. In these systems, the variations in the U/W ratio affect their electronic structure and physical properties. A large U/W ratio inhibits charge fluctuations, eventually producing a metal-insulator transition. The metal-insulator transition can be controlled by changing either the bandwidth or the band filling [1].

The $Ca_{1-x}Sr_xVO_3$ series allows controlling the one electron bandwidth at a fixed ($3d^1$) band filling [1]. The bandwidth is changed due to a distortion caused by the different ionic radius of the $Sr^{2+}$ and $Ca^{2+}$ ions. In particular, $SrVO_3$ has a cubic perovskite structure with a V–O–V angle of around 180º, whereas the $CaVO_3$ compound is orthorhombic with a V–O–V angle of about 160º. The orthorhombic distortion decreases the inter-cluster hopping thus reducing the one electron bandwidth.

Early photoemission spectra of $Ca_{1-x}Sr_xVO_3$ show a *coherent* peak and an *incoherent* structure [2,3]. The spectra present a systematic transfer of spectral weight from the *coherent* to the *incoherent* peak in the series. The transitions from the coherent and incoherent features are also reflected in optical measurements [4,5]. The changes in the effective mass m* obtained from the optical measurements were large [5], although the effective mass m* deduced from the electronic specific heat did not change much [6].

The *coherent* structure in the spectra can be related to the calculated V 3d density of states [2,3]. But the *incoherent* feature reflects a distinct many body effect within the photoemission process. DMFT calculations attribute the *coherent* structure to the quasi-particle peak, and the *incoherent* feature to the remnant of the lower Hubbard band [7,8]. Recent studies focused on the higher correlation at the surface [9], as well as on the energy dependence of the photoemission spectra [10-12].

In this work, we studied the electronic structure of the $Ca_{1-x}Sr_xVO_3$ series using a cluster model. The calculation includes a non-local *coherent screening* to account for the metallic charge fluctuations. Both the *coherent* and *incoherent* structures are *well-screened* states, while the *poorly-screened* state appears only at deeper energies. The main parameter is the non-local transfer integral T* which decreases with the lattice distortion. This produces the observed transfer of spectral weight from the *coherent* to the *incoherent* structure.

The cluster considered in this work was a $V^{4+}$ ion surrounded by an $O^{2-}$ octahedra. The cluster model was solved using the standard configuration interaction method. The ground state was expanded in the $3d^1$, $3d^2\underline{L}$, $3d^3\underline{L}^2$, etc. configurations, where $\underline{L}$ denotes a hole in the O 2p orbitals. The main parameters of the model were the charge-transfer energy $\Delta$, the Mott-Hubbard energy U and the p-d transfer integral $T_\sigma$. The removal state was obtained by removing an electron from the ground state. Finally, the photoemission spectra was calculated using the sudden approximation. The main parameters of the cluster model were $\Delta = 2.8$ eV, $U = 5$ eV, and $T_\sigma = 3.3$ eV for $SrVO_3$. The values of $\Delta$ and U remain constant, but the value of $T_\sigma$ decreases to 2.8 eV in $CaVO_3$. The crystal field splitting was $10Dq = 2.1$ eV and the p-p transfer integral was $pp\pi-pp\sigma = 1.3$ eV. These values of the parameters gave the best results and are in good agreement with previous reports [13].

Besides the usual local *ligand screening*, the calculation included a non-local *coherent screening*. This comes from a extended state at the Fermi level and is related to the metallic charge fluctuations. The ground state was therefore expanded in the $3d^1$, $3d^2\underline{L}$, $3d^2\underline{C}$, $3d^3\underline{L}^2$, $3d^3\underline{CL}$, $3d^3\underline{C}^2$, etc. configurations, were $\underline{C}$ denotes a hole in the *coherent* band. This process was recently included in spectroscopic studies of $V_2O_3$ [14] and $VO_2$ [15], following the original ideas proposed by Kotani and Toyozawa [16] and Bocquet *et al.* [17]. The extra parameters, in this case, are the charge transfer $\Delta^*$ and the transfer integral $T^*$ [14]. The value of $\Delta^*$ is roughly related to the width of the occupied part of the one-electron V 3d band. On the other hand, the transfer integral $T^*$ is related to second order V-O-V hopping processes. The values of both $\Delta^*$ and $T^*$ are reduced as the V-O-V angle decreases in the series (the value of $T_\sigma$ is also reduced due to the distortion, although the reduction is smaller than for $T^*$).

The ground state of $SrVO_3$ is formed mainly by the $3d^1$ (27%), $3d^2\underline{L}$ (48%), and $3d^2\underline{C}$ (12%) configurations. The ground state of $CaVO_3$ is formed mostly by the $3d^1$ (29%), $3d^2\underline{L}$ (47%), and $3d^2\underline{C}$ (12%) configurations. The covalent mixture between the V 3d and O 2p states ($3d^2\underline{L}$) is rather large because $\Delta$ is small and $T_\sigma$ is large. The *coherent* contribution to the ground state ($3d^2\underline{C}$) is smaller because, although $\Delta^*$ is small, $T^*$ is much smaller.

Figure 1 shows the removal spectra of $SrVO_3$ and $CaVO_3$ projected on the main final state configurations. For $SrVO_3$, the *coherent* structure about –0.4 eV is formed mainly by the $3d^1\underline{C}$ configuration (24%), whereas the *incoherent* feature around –1.6 eV is formed mostly by the $3d^1\underline{L}$ configuration (35%). The lower Hubbard band appears only at –7.2 eV and is formed mainly by the $3d^0$ configuration (55%).

The value of U dictates the position of the $3d^0$ configuration before the hybridization is switched on. In turn, the values of $\Delta$ and $\Delta^*$ are related to the initial positions of the $3d^1\underline{L}$ and $3d^1\underline{C}$ configurations. The $T_\sigma$ ($T^*$) hybridization pushes the $3d^1\underline{L}$ ($3d^1\underline{C}$) configuration towards the Fermi level, while the $3d^0$ configuration is pushed to larger binding energies. Finally, the $T_\sigma$ ($T^*$) hybridization is also related to the relative intensity of the $3d^1\underline{L}$ ($3d^1\underline{C}$) configuration.

The distribution of the spectral weight in $CaVO_3$ is similar to that in $SrVO_3$, as in Fig. 1. The *coherent* structure about –0.4 eV is formed mainly by the $3d^1\underline{C}$ configuration (31%), whereas the *incoherent* feature around –1.6 eV is formed mostly by the $3d^1\underline{L}$ configuration (442%). This is in agreement with resonant photoemission spectra of $CaVO_3$ [18], which show a resonant enhancement of the feature about –1.6 eV. Finally, the lower Hubbard band around –6.4 eV is formed mainly by the $3d^0$ configuration (56%).

The results above show that both the *coherent* and *incoherent* features correspond to *well screened* states. The screening charge of the *incoherent* structure comes from the local *ligand channel* ($3d^1\underline{L}$), whereas the *coherent* feature comes from the non-local *coherent channel* ($3d^1\underline{C}$). On the other hand, the *poorly screened* state ($3d^0$) corresponds to the lower Hubbard band. The system is in the charge transfer regime, $\Delta$ ($\Delta^*$) < U, with the $3d^1\underline{L}$ ($3d^1\underline{C}$) states close to the Fermi level [19]. Finally, the lowest energy (metallic) charge fluctuations involve transitions from $3d^1\underline{C}$ to $3d^2$ states.

The *coherent* structure, mostly $3d^1\underline{C}$, has an almost pure V 3d character, whereas the *incoherent* feature, mainly $3d^1\underline{L}$, is of mixed O 2p – V 3d character. Indeed, the $3d^1\underline{C}$ final state can only be obtained from the $3d^2\underline{C}$ ground state by removing a V 3d electron. But the $3d^1\underline{L}$ final state can be reached by removing a V 3d electron (from $3d^2\underline{L}$) or an O 2p electron (from $3d^1$). Figure 2 shows the relative weight of the V 3d and O 2p contributions to the removal spectra. The relative cross sections of the V 3d and O 2p levels were adjusted to a photon energy of 21.2 eV [20]. These results show that at low photon energy the *incoherent* structure contains considerable O 2p weight. But, the relative cross section increases the V 3d contribution for higher photon energies, see below.

Figure 3 shows the *coherent* and *incoherent* removal spectra along the $Ca_{1-x}Sr_xVO_3$ series. The $T^*$ parameter changes from 0.27 to 0.22 eV, $\Delta^*$ from 0.75 to 0.55 eV, and $T_\sigma$ from 3.3 to 2.8 eV. The calculations are in qualitative agreement with the experimental spectra taken from Ref. 2. In particular, the results reproduce the transfer of spectral weight from the *coherent* to the *incoherent* region. This transfer of spectral weight takes place through a non-trivial second order process. The spectral weight goes from the $3d^1\underline{C}$ to the $3d^0$ configuration via the $T^*$ hybridization, and then back to the $3d^1\underline{L}$ configuration through the $T_\sigma$ transfer integral (there is not any single particle matrix element between the $3d^1\underline{L}$ and the $3d^1\underline{C}$ configurations). The crucial parameter here is the transfer integral $T^*$ of the *non-local screening channel*. The decrease of the V-O-V angle in the series reduces the effective inter-cluster transfer integral $T^*$. This causes the decrease of the *coherent* peak and the transfer of spectral weight to the *incoherent* peak. A further reduction of $T^*$, below a critical value, would suppress the metallic charge fluctuations. This effect is responsible for the metal-insulator transition in the $YTiO_3$–$LaTiO_3$–$SrVO_3$–$VO_2$ series [21]. The lattice distortion in $Ca_{1-x}Sr_xVO_3$ is not large enough to reach the critical value, and therefore the otherwise expected bandwidth controlled metal-insulator transition is not realized.

There is a controversy in the interpretation of the low energy photoemission spectra of $Ca_{1-x}Sr_xVO_3$. This concerns the relative bulk or surface character of the *coherent* and *incoherent* structures. The relative correlation effects at the surface are indeed enhanced due to the reduced dispersion [9]. But the variation in the spectra cannot be explained by changes in the surface sensitivity, because the spectra were obtained using the same photon energy and thus the same sampling depth. A more drastic variation is observed in the spectra of the $YTiO_3$–$LaTiO_3$–$SrVO_3$–$VO_2$ series, where the weight of the *coherent* structure is completely suppressed in the insulating phase [21]. The same effect is observed in the insulating phase of $VO_2$ where the *coherent* feature also disappears [15]. The suppression of the *coherent* structure, in these cases, cannot be explained in terms of surface effects.

Recent energy dependent photoemission studies showed an increase of the *coherent* peak [10-11]. This enhancement was attributed to a relatively larger bulk character of the *coherent* structure, because a higher energy photoelectron has a relatively larger escape depth. Figure 4 shows the calculated photoemission spectra at 21.2 eV (UPS) and 1486.6 eV (XPS). The calculated results are in good agreement with the experimental UPS and XPS spectra taken from Ref. 10. Thus, the changes in the spectra can be mostly attributed to the energy dependence of the cross section. The small discrepancy between the calculation and the experiment is attributed to residual surface effects.

There is an apparent discrepancy in the physical properties of the $Ca_{1-x}Sr_xVO_3$ material. The changes in the excitation spectra are larger than in the thermodynamics properties. This difference can be reconciled taking into account the relative weight of the *coherent* contribution. The *coherent* contribution to the ground states ($3d^2\underline{C}$) in $SrVO_3$ (12%) and $CaVO_3$ (12%) are similar. But the *coherent* structure in the removal spectra ($3d^1\underline{C}$) is 25% larger for $SrVO_3$ than for $CaVO_3$. This suggests that thermodynamics properties should change little in the $Ca_{1-x}Sr_xVO_3$ series; whereas the excitation spectra, like photoemission and optical spectra, should present larger changes. This helps to explain why the effective mass m* obtained from the electronic specific heat do not change [6]; whereas the effective mass m* deduced from the optical response decreases 25% from $SrVO_3$ to $CaVO_3$ [5] (the changes in the optical m* are in agreement with the observed decrease in the *coherent* structure above). The above results show that the different physical properties should be compared with caution.

In conclusion, we studied the electronic structure of the $Ca_{1-x}Sr_xVO_3$ series using a cluster model. Both the *coherent* and *incoherent* structures in photoemission correspond to *well screened* states. The *coherent* structure ($3d^1\underline{C}$) is mostly due to a non-local *coherent screening*, whereas the *incoherent* feature ($3d^1\underline{L}$) is mainly related to the local *ligand screening*.

The *poorly screened* state ($3d^0$) forms the lower Hubbard band and appears deeper in energy. The increased distortion in the series decrease the V-O-V angle and thus the non-local transfer integral T*. This produces the observed transfer of spectral weight from the *coherent* to the *incoherent* structure. A further decrease of T* would eventually induce a bandwidth controlled metal-insulator transition.

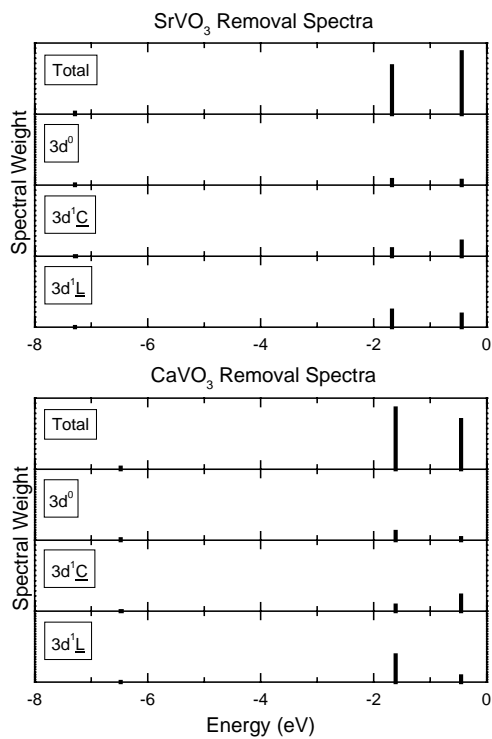

**Figure 1:** Removal spectra of $SrVO_3$ and $CaVO_3$ projected on the main final state configurations.

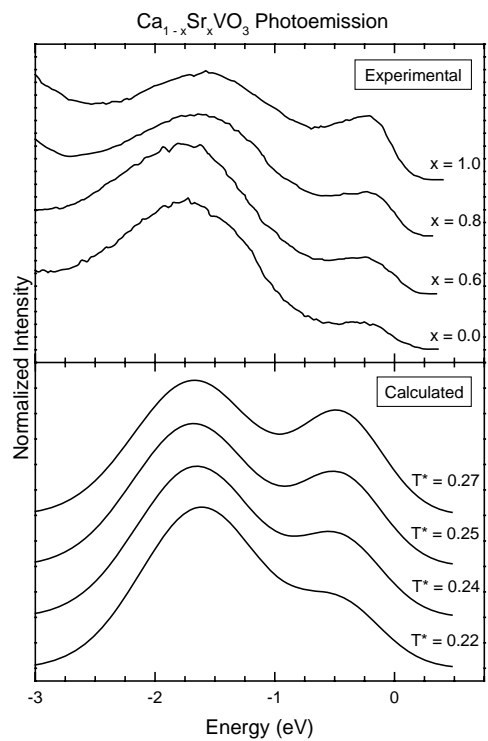

**Figure 3:** Experimental and calculated photoemission spectra of the $Ca_{1-x}Sr_xVO_3$ series taken from Ref. 2.

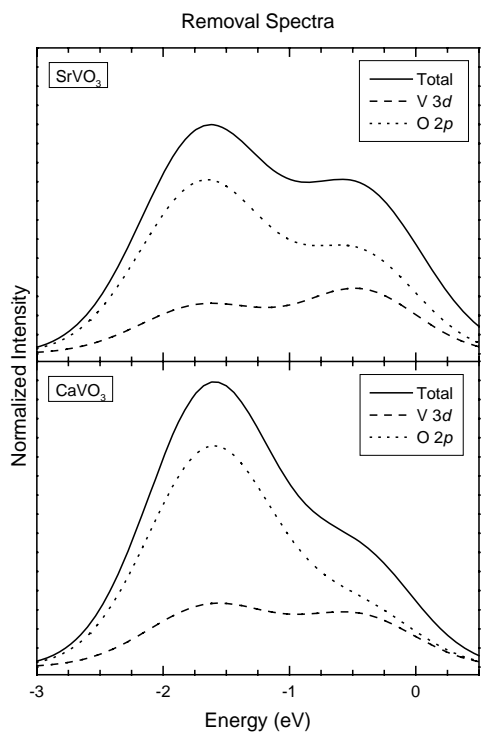

**Figure 2:** Removal spectra of $SrVO_3$ and $CaVO_3$ projected on the V 3d and O 2p contributions.

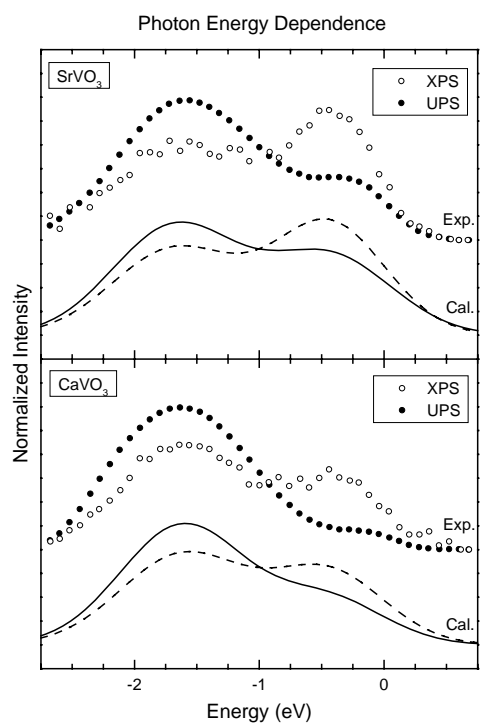

**Figure 4:** Experimental and calculated UPS and XPS spectra of the $Ca_{1-x}Sr_xVO_3$ series taken from Ref. 13.